\documentclass{tlp}

\usepackage{amsmath}

\usepackage{listings}

\usepackage{url}

\makeatletter
\renewcommand\section{%
  \@startsection {section}{1}{\z@}
    {-19.5\p@ \@plus -6.5\p@ \@minus -3.25\p@}
    {6.5\p@ \@plus \z@ \@minus 1\p@}
    {\normalfont\normalsize\bfseries}%
}
\renewcommand\subsection{%
  \@startsection{subsection}{2}{\z@}
    {-19.5\p@ \@plus -3.25\p@ \@minus -3.25\p@}
    {6.5\p@ \@plus \z@ \@minus 1\p@}
    {\normalfont\normalsize\bfseries\itshape}%
}
\makeatother
\newcommand*{\guideline}[1]{\subsection{#1}}

\newcommand*{\recommend}[1]{\paragraph{\textbf{If in doubt:}} #1}

\newcommand*{\Cplusplus}{{C\nolinebreak[4]\hspace{-.05em}\raisebox{.4ex}
{\tiny\bf ++}}}

\begin{document}

\title{Coding Guidelines for Prolog}

\author[M. A. Covington et al.]
       {MICHAEL A. COVINGTON \\
       Institute for Artificial Intelligence,
       The University of Georgia,
       Athens, Georgia,
       U.S.A. \authorbreak
       \email{mc@uga.edu}
       \and
       ROBERTO BAGNARA\thanks{The work of R.~Bagnara has been partly supported
         by PRIN project
         ``AIDA2007 --- Abstract Interpretation Design and Applications.''} \\
       Department of Mathematics,
       University of Parma,
       Italy \\
       BUGSENG srl,
       Parma,
       Italy \authorbreak
       \email{bagnara@cs.unipr.it}
       \and
       RICHARD A. O'KEEFE \\
       Department of Computer Science,
       University of Otago,
       Dunedin,
       New Zealand \authorbreak
       \email{ok@cs.otago.ac.nz}
       \and
       JAN WIELEMAKER \\
       Department of Computer Science,
       VU University Amsterdam,
       The Netherlands \authorbreak
       \email{j.wielemaker@cs.vu.nl}
       \and
       SIMON PRICE \\
       Intelligent Systems Laboratory,
       University of Bristol,
       United Kingdom \authorbreak
       \email{simon.price@bristol.ac.uk}
}

\submitted{November 15, 2009}
\revised{November 11, 2010}
\accepted{May 2, 2011}

\maketitle

\begin{abstract}
Coding standards and good practices are fundamental to a disciplined
approach to software projects, whatever programming languages they employ.
Prolog programming can benefit from such an approach, perhaps more
than programming in other languages. Despite this, no widely accepted
standards and practices seem to have emerged up to now.
The present paper is a first step towards filling this void:
it provides immediate guidelines for code layout, naming conventions,
documentation, proper use of Prolog features, program development,
debugging and testing.  Presented with each guideline is its rationale
and, where sensible options exist, illustrations of the relative pros
and cons for each alternative.
A coding standard should always be selected on a per-project basis,
based on a host of issues pertinent to any given programming project;
for this reason the paper goes beyond the mere provision of normative
guidelines by discussing key factors and important criteria that
should be taken into account when deciding on a fully-fledged coding
standard for the project.
\hfill\\
\hfill
\end{abstract}

%=======================================================================
\section{Introduction}
\label{sec:introduction}
%=======================================================================

The purpose of programming languages is to make programs readable by people.
Coding standards enhance this function,
especially when multiple programmers and a need for maintainability are present,
but also even in the small projects of a single programmer
(one must, after all, read and debug one's own work).
Coding standards for Prolog are particularly needed for several
reasons:
\begin{description}
\item[Availability.]
As far as we know, a coherent and reasonably complete set of coding
guidelines for Prolog has never been published.  Moreover, when we look
at the corpus of published Prolog programs, we do not see a de facto
standard emerging. The most important reason behind this apparent
omission is that the small Prolog community, due to the lack
of a comprehensive language standard, is further fragmented into
sub-communities centered around individual Prolog systems, none of
which has a dominant position (contrast this with the situation of
Java and Sun's coding conventions for that language
\cite{SunMicrosystems99} or the precedents set for C by the UNIX
source code).
\item[Language.]
Language features that contribute to the power of the language make it
quite easy ---especially for the non-expert--- to get things wrong,
sometimes in ways that are difficult to diagnose. For example, the
lack of prescriptive typing contributes to the suitability of Prolog
for quick prototyping in some application domains. At the same time,
the fact that no redundant type information is available to the
development tools makes discipline particularly important.
Besides \emph{types}, Prolog developers and maintainers are confronted
with \emph{modes}: the same arguments of procedures can be inputs, outputs,
or both.  This gives Prolog
conciseness and elegance through reversible predicates but makes it necessary
to keep
track of which modes are supposed to be supported (within reasonable
computational complexity limits) by which predicates.
\item[Compiler technology.]
Most Prolog compilers give no warnings except about singleton variables.
Even though more advanced tools and development environments exist, a
disciplined approach is still (and will always be) the best device available
to those programming in standard Prolog (and any other language).
\end{description}

It is important to stress that we are not announcing \emph{the} coding
standard for Prolog.  This is a paper with five authors and, on some
points, more than five opinions.
Rather, we address numerous
issues that the maker of a full-fledged coding standard will have to confront.
In some cases, there are good reasons to prefer one alternative to another.
Other decisions are arbitrary, like driving on the left or on the right;
a community of programmers can choose any of several ways of doing something
as long as they are consistent.
The classic style guide is {\it The Elements of Programming Style}
\cite{KernighanP78}. Published in 1978 and using examples in
PL/I and Fortran, the majority of its advice is independent of programming
language and still much needed.
Indeed, some of our own guidelines are not Prolog-specific but are included
because no style guide for practitioners would be complete without them.

The amount of standardization needed depends on the scale and duration
of the programming project.
Just like a novel or a scientific paper, the quality of a computer
program can only be judged from the degree to which it satisfies the
programmer's objectives,
i.e., ``it works'': for example, we might say that a novel works
if it becomes a best-seller, the manual of a device works if it enables
the average user to effectively operate the device, a scientific paper
works if it is published in a prestigious journal.  Computer programs
written for different purposes may or may not ``work'' in different ways.
Only for very simple programs written to solve very simple
tasks we can say that \emph{the program works if it functions correctly}.
For even moderately complex programs a more sensible definition could be:
\emph{the program works if its developers and maintainers are able
to approach the expected behavior of the program over its intended
lifespan.}

In some cases correctness and maintainability over a long
period of time are not important.  For instance, in rapid prototyping
of applications (something for which Prolog has its advantages) we might
say that \emph{the program (prototype) works if (despite its errors
and limitations) it demonstrates that the approach is feasible.}
Still, the prototype will be incomplete until it is finished, and its
chances of being finished one day (as opposed to collapsing and being
abandoned before being of any use) depend on qualities such as
readability, extensibility, and whatever else helps the development team.

In a sense, the text of a program is not very
different from an argumentative essay.  Like an argumentative essay,
a nontrivial program is addressed to an \emph{audience}, which needs
to be accurately identified for the argumentation part of the program
to be effective.  As for any argumentation, different audiences will
require and be prepared to share a different set of \emph{premises}
(hypotheses and preconditions explicitly mentioned in the program text)
and \emph{assumptions} (hypotheses and knowledge that are left implicit
but that still are essential for the comprehension of the program).
Not to mention that, in programming, the intended audience is also
heavily influenced by the tools it is used to.

To summarize, the existence of very different purposes and wildly
different audiences are such that a full-fledged coding standard
can only be decided upon on a per-project basis.  Moreover, different,
equally reasonable and effective coding standards can, on specific
points, recommend plainly opposite things: a coherent whole
matters more than the individual bits.

In this paper, which evolved from \cite{Covington02}, we
introduce a set of coding guidelines that can serve as a starting
point for the development of effective coding standards. We highlight
the aspects of Prolog program development that deserve particular
attention and can benefit from regulation; we illustrate
the rationale that is behind each of the proposed guidelines; when
alternative guidelines can achieve the desired effect, their relative
merits are discussed.

The paper is organized as follows: in
Sections~\ref{sec:layout}, \ref{sec:names} and~\ref{sec:documentation}
we discuss guidelines concerning code layout, the naming of program
entities, and documentation, respectively;
Section~\ref{sec:language-idioms} concerns the effective use of
language features;
Section~\ref{sec:development} deals with the
development, debugging and testing of program units and their
interfaces;
Section~\ref{sec:conclusion} concludes.

%=======================================================================
\section{Layout}
\label{sec:layout}
%=======================================================================

Do not let your programs be hampered by inconsistent layout.  Poor layout
is painful to work with, is distracting, and can lead to otherwise perfectly
avoidable mistakes.  Moreover, most text editors, if properly operated,
can provide significant assistance in ensuring layout consistency.

%-----------------------------------------------------------------------
\guideline{Indent with spaces instead of tabs.}

You cannot normally predict how tabs will be set in the editors and
browsers others will use on your code.  Moreover, mixing tabs and
spaces makes searches and substitutions more problematic.  All in all,
it is better not to use tabs altogether: almost any editor can be set to
substitute spaces for tabs automatically when saving your file.

%-----------------------------------------------------------------------
\guideline{Indent consistently.}

Code is more readable if it is indented consistently.
Even though the choice of the ``right'' indent size often leads
to friction within the development team, a decision has to be made.
Most editors default to an indent size of 8 and while such a large
indent does make shallowly nested indentation obvious, it uses up
excessive horizontal screen real estate and makes even moderately nested
indentation hard to read. A small indent size of 2 spaces is popular
with some experienced Prolog programmers but can make it hard for less
keen eyed readers or for those using variable pitched fonts to
interpret the indentation depth.  The compromise value of 4 has proven
popular and practical in both Prolog and other well known languages.

\recommend{Use an indent size of 4 spaces.}

%-----------------------------------------------------------------------
\guideline{Limit the length of source code lines.}

Lines longer than 80 characters are almost always difficult to read.
While 80-column-wide screens are no longer in widespread use, 80 columns is still the
default width for many text editors.
Some editors may also display the
last column in an inconvenient way, so limiting the line length to 79
or 78 characters is a good idea.

High-quality printed listings can usually accommodate no more than 65 or 70
characters per line, depending on the type size and paper size.
For maximum readability you may want
to lower this limit down to 55 characters \cite{Covington94}.
%-----------------------------------------------------------------------
\guideline{Limit the length (number of lines) of clauses.}

The ideal situation would of course be for each clause to fit onto
an ordinary computer screen.  So 24 lines is the safest choice but
lengths up to 48 can fit on modern computer displays.  You can consider
this a limit not to be crossed, the exceptions being for predicates doing
long but conceptually simple sequences of I/O or graphic operations.

More generally, always consider simplifying long clauses, whether or not
they fit onto your screen. Since in Prolog the only way to have a loop
is to introduce another predicate,\footnote{There have been several
proposals to provide in-line loop constructs. See, for example,
``Logical Loops'' \cite{Schimpf02}, ``Declarative Loops and List
Comprehensions for Prolog - B-Prolog''
(\url{www.probp.com/download/loops.pdf}), or Lambdas in ISO Prolog
(\url{http://www.complang.tuwien.ac.at/ulrich/Prolog-inedit/ISO-Hiord.html}).
We do not discuss these due to lack of agreement in the community.}
and case analyses often (though not always) deserve their own predicates
too, it is no hardship to follow this guideline.

%-----------------------------------------------------------------------
\guideline{Be consistent in the use of space around commas.}

The simplest approach is to follow each comma with a space,
just as in English.
This improves readability by increasing the visual separation among
different arguments of a term.  Moreover, as this is consistent with
the typesetting conventions used for many other
languages,\footnote{Both natural and formal languages, such as Ada~95
\cite{AusnitHoodJPO97} and Java \cite{SunMicrosystems99}.}
doing so often avoids a disturbing mismatch between the program text,
its comments, and the string literals used in the program.

The simplicity of the above guideline is valuable, but it misses the
opportunity to reduce the decoding effort for readers.
An alternative is thus to use space in a way that helps in recognizing the
various uses of commas.\footnote{It was pointed out long ago by critics
of Prolog that different commas mean different things.  Of course, Prolog
is by no means alone in this: take C as an example.}
One possibility is to distinguish between:
\begin{itemize}
\item
comma meaning \emph{and-then}: follow it by one or more spaces, an optional
comment and a newline (see also
Guideline~\ref{guideline:put_each_subgoal_on_a_separate_line});
\item
comma separating arguments of a goal: follow it by one or more spaces;
\item
comma separating elements of a data structure: do not follow it by spaces.
\end{itemize}
Of course, in order to break long lines, extra newlines should be allowed
in the second and third case.
In addition, it is tempting to use the ability of putting one or more
spaces to line stuff up, as in:
\begin{verbatim}
shorten(person(Name,       Address, Age),
        person(Short_Name, Address, Age)) :-
    ...
\end{verbatim}
Similarly, it may seem natural (to one person, on a particular day,
and in a particular context) to relax the rules in order to help
distinguishing ``levels'', as in \verb"[[1,2,3], [4,5], [6,7,8,9]]".
Note also that \verb"[alpha, beta, gamma]" looks nice because the terms
are longer than in \verb"[1,2,3]", \emph{but} if you allow that to make
a difference, you can no  longer rely on spaces to tell you anything
you want to know.

This is why layout remains an art rather than a science.
There will always be a friction between the rules ---which tend to be
either too complex or too rigid but allow for uniformity--- and taste
---which can adapt to particular situations in often nicer ways but
is personal, variable and sometimes inconsistent.

For small individual projects, following some general guidelines with
flexibility and applying taste on a case-by-case basis can give good
results.  However, for projects that are bigger and/or have several
developers it is best to take style decisions on the basis of explicit
and precise principles rather than on the basis of what feels good
at the time: sticking to precise rules may occasionally lead to some
program fragments that do not look very good, but the overall
readability of the project as a whole will benefit.
%-----------------------------------------------------------------------
\guideline{Begin each clause on a new line and indent all but the first line of each clause.}

For example, write:
\begin{verbatim}
same_length([], []).
same_length([_|L1], [_|L2]) :-
    same_length(L1, L2).
\end{verbatim}
%-----------------------------------------------------------------------

\guideline{Put each subgoal on a separate line.}
\label{guideline:put_each_subgoal_on_a_separate_line}

Putting each subgoal on a new line greatly enhances readability.
So, no matter how short some
subgoals may be, it is better to use one line for each of them.
For example:
\begin{verbatim}
ord_union_all(N, Sets0, Union, Sets) :-
    A is N / 2,
    Z is N - A,
    ord_union_all(A, Sets0, X, Sets1),
    ord_union_all(Z, Sets1, Y, Sets),
    ord_union(X, Y, Union).
\end{verbatim}
The only exception might be for short sequences of closely related subgoals,
such as those involving \verb"write" and \verb"nl".
For instance, one could have
\begin{verbatim}
    write('CPU time = '), write(T), write(' msec'), nl.
\end{verbatim}
Note though that the \verb"format/[1,2,3]" built-in predicates
(not yet in the Prolog standard, but provided by quite a few Prolog systems
with reasonably compatible implementations) provide a better way
to underscore that, conceptually, several output operations constitute
a single step.
%-----------------------------------------------------------------------
\guideline{Use vertical space consistently to improve readability.}

It is natural for vertical distance to reflect the
logical distance.  For example: begin the next clause of the same predicate
on the next line;  skip a line before the first clause of the next related
predicate;  skip two lines before the first clause of the next unrelated
predicate.
Notice that the unindented clause head is enough to separate clauses
within the definition of a single predicate: no lines need be skipped here.
%-----------------------------------------------------------------------
\guideline{Comment source files, not just the predicates within them.}

In addition to the comments on particular clauses or predicates,
a Prolog source file should begin with some sort of standard header
with meta-data about the file: name, version, author(s), revision history,
copyright, license and other information.
Next there should be the module declaration (in case a module system is
used) and maybe other declarations.
Immediately after the declaration a detailed comment should
explain what the file is all about, what the central ideas are,
possibly with illustrations of the data structures used.

It is best if this long explanatory comment is readable as plain text with a minimum
of distracting punctuation: it is thus better to use plain \verb"/*"
and \verb"*/" delimiters with no extra stars.\footnote{But possibly with
dashes at the top and bottom to increase visual
separation.}  (The \verb"%"
comment delimiter is reserved for comments pertaining to individual predicates, as
described below.)
If a file divides naturally into several sections, each section can
have its own such comment.
%-----------------------------------------------------------------------
\guideline{Use layout to make comments more readable.}

While text editors can provide significant help in properly formatting code,
for comments the developer is usually on his or her own.  Nonetheless, a little
extra care can do much to increase the readability and usefulness of comments.
For example, instead of writing:
\begin{verbatim}
% This predicate classifies C as whitespace (ASCII < 33), alphabetic
% (a-z, A-Z), numeric (0-9), or symbolic (all other characters).
\end{verbatim}
write something like this:
\begin{verbatim}
% This predicate classifies C as:
%   - whitespace (ASCII < 33);
%   - alphabetic (a-z, A-Z);
%   - numeric (0-9); or
%   - symbolic (all other characters).
\end{verbatim}
Indented lists like this are much easier to read than lists disguised
as paragraphs.  It is also much less work to add or remove an item or
add comments about an item.
%-----------------------------------------------------------------------
\guideline{Avoid comments to the right of the code, unless they are
inseparable from the lines on which they appear.}

Comments on the right are problematic, since they so easily get out of
step with the code when small changes are made to the program.
Moreover, as soon as there is more than one
end-of-line comment in the same predicate, it looks bad if they do not
begin in the same column, and maintaining this alignment when editing
the program becomes an extra chore.

If you cannot dispense with comments to the right of the code,
at least make them very short: just a few crucial hints and,
if needed, a pointer to the predicate documentation.
For example:
\begin{verbatim}
%% pred(...)
%  ...
%  [Note 1] Long comment about how pred/n depends on subgoal/k...
%  ...

pred(...) :-
    ...
    subgoal(...),        % See [Note 1] above.
    ...

\end{verbatim}

Note that it is perfectly legitimate to write single- and multiple-line
comments on their own lines, indented like the other lines in a clause
body.

%-----------------------------------------------------------------------
\guideline{Consider using comments consistently as reminders.}

With some discipline you can effectively mark the points that require
your attention at a later time.  One possibility is to use the
\verb"%" comment delimiter to create ``tags'' that can then easily
and reliably retrieved with string search.  For example:

\begin{verbatim}
%TBD: <short hint on what remains to be done>
%FIXME: <what/why this does not work in all cases>
%HACK: <why this is not completely satisfactory>
\end{verbatim}
Here \verb"TBD" abbreviates \emph{to be done.}
These abbreviations correspond to the \verb"//TODO:" comments familiar to C\#
and Java programmers in various development environments.

In more extreme cases, you can guarantee that an unfinished section of
your program is not used by inserting Prolog code that will not work.
In an unfinished predicate, insert a call to something like
\verb"tbd('description of what is to be done')"
and do not define the predicate \verb"tbd".
Then any attempt to call the predicate will throw an error.

A related tip is to always write a special comment, like \verb"%DEBUG",
at the end of each line added solely for debugging purposes.
Indeed, forgetting to delete debugging code is a common mistake.
By tagging that code with \verb"%DEBUG" or similar, it becomes easy
to search and delete all the temporary debugging goals and clauses.

%-----------------------------------------------------------------------
\guideline{Make clauses understandable in isolation.}

As far as you can easily do so, enable the reader to understand each clause
without having read everything above it.
For example, if a test is unnecessary because a cut guarantees its truth,
say so in a comment at the appropriate place.
For example, write:
\begin{verbatim}
remove_duplicates([First|Rest], Result) :-
    member(First, Rest),
    !,
    remove_duplicates(Rest, Result).
remove_duplicates([First|Rest], [First|New_Rest]) :-
    % First does not occur in Rest.
    remove_duplicates(Rest, New_Rest).
\end{verbatim}
The comment keeps the clause from being misunderstood if the
reader has not read the preceding clause.
%-----------------------------------------------------------------------

\guideline{Indent an additional level between {\tt repeat} and the corresponding cut.}

This makes a \verb"repeat" structure look more like a loop. Here is an example:
\begin{verbatim}
process_queries :-
    repeat,
        read_query(Q),
        handle(Q),
        Q = [quit],
    !,
    write('All done.'), nl.
\end{verbatim}

%-----------------------------------------------------------------------
\guideline{Decide how you want to break long clause heads and subgoals.}

Even if you try to limit the number of arguments of your predicates
(see Guideline~\ref{guideline:try_to_limit_the_number_of_predicate_argument}),
and, or perhaps because, you choose your names wisely
(see Guidelines~\ref{guideline:first_guideline_on_choosing_names}%
--\ref{guideline:last_guideline_on_choosing_names}),
you will occasionally end up with a clause head or a subgoal that is too
long to fit on a single line.
In this case, two alternatives make sense:
\begin{verbatim}
long_predicate_name(long_arg1, long_arg2, long_arg3,
                    long_arg4, long_arg5) :-
    ...
\end{verbatim}
and
\begin{verbatim}
long_predicate_name(
    long_arg1, long_arg2, long_arg3,
    long_arg4, long_arg5
) :-
    ...
\end{verbatim}
The second alternative may be perceived as less pretty, but it keeps
predicate name length from affecting indentation.
By contrast, the first alternative forces you to change the indentation
of \verb"long_arg4" every time the predicate is renamed
(which happens fairly often during development).
However, the lengths of the arguments still affect layout.
Alternatives include placing each argument on its own line
or placing only semantically related groups of arguments on the same line.
%-----------------------------------------------------------------------

\guideline{Decide how to format disjunctions and if-then-elses.}

Prolog programmers are deeply divided on how to format the semicolon
(which means ``or'') and the if-then-else structure (\verb"Goal1 -> Goal2 ; Goal3").
Above all, note that:
\begin{itemize}
\item
Heavy use of semicolons and if-then-elses may indicate that the
program is not broken up into clauses correctly.  The main ways to
indicate disjunction and conditionality are to provide multiple
clauses and control the success conditions of each clause.
Use semicolons to mark small portions of a clause as nondeterministic
alternatives.
Use if-then-else for deterministic conditionals, as an alternative to cuts.
\item
A semicolon at the end of a line can easily go unnoticed and result
into a serious programming error.
\item
Programmers do not remember the relative precedence of commas and
semicolons; \verb"a,b;c" means \verb"(a,b);c" but this fact is not
obvious.  Thus, layout and parentheses to indicate precedence are
recommended.
\end{itemize}
All but the simplest disjunctions should be displayed prominently using layouts
such as
\begin{verbatim}
    (
        disjunct_1
    ;
        disjunct_2
    ;
        disjunct_3
    )
\end{verbatim}
or the more compact
\begin{verbatim}
    (   disjunct_1
    ;   disjunct_2
    ;   disjunct_3
    )
\end{verbatim}
In both styles, parentheses are always present, the closing parenthesis
is exactly below the opening one, and the semicolons stick out in a way
that makes misunderstanding much more difficult.
If all of this seems too prolix, recall the first bulleted item above.

The second style, besides compactness, has the advantage that it
naturally generalizes to if-then-elses as follows:
\begin{verbatim}
    (   test_1 ->
        test_1_is_true
    ;   test_2 ->
        test_2_is_true_and_test_1_is_false
    ;   both_are_false
    )
\end{verbatim}
A variation that lets the `\verb+->+' operators be more visible at the
expense of more vertical space usage is:
\begin{verbatim}
    (
        test_1
     ->
        test_1_is_true
     ;
        test_2
     ->
        test_2_is_true_and_test_1_is_false
     ;
        both_are_false
    )
\end{verbatim}
This can also be written as
\begin{verbatim}
    (test_1 ->
        test_1_is_true
    ;
        (test_2 ->
            test_2_is_true_and_test_1_is_false
        ;
            both_are_false
        )
    )
\end{verbatim}
which is more expensive both in terms of horizontal and vertical space,
but some consider it more readable.
A third alternative is to put \verb"->" and \verb";" at the beginnings of lines:
\begin{verbatim}
    (test_1
    ->  test_1_is_true
    ;   (test_2
        ->  test_2_is_true_and_test_1_is_false
        ;   both_are_false
        )
    )
\end{verbatim}
A further variation is to put each \verb"->" and \verb";" on a line by itself.
Whichever style you choose, you should work out how it will handle nesting.

%-----------------------------------------------------------------------
\guideline{Consider using automated tools to improve readability.}

A \emph{pretty-printer} is a program that prints your programs in a neat,
readable form.  Pretty-printers are sometimes called \emph{code beautifiers}.

To be more precise, we should recall that,
historically, a ``pretty-printer'' was a program that altered the white
space in the source code (adding and removing spaces and line breaks)
in order to achieve consistent indentation.  For C, \verb"cb" and
GNU's \verb"indent" are programs of this type.
Some Lisp and Scheme systems have a \verb"pp" function for printing
functions prettily.
Pretty-printing of this type can greatly improve code readability.

Automatic (re-)indentation is built into many editors,
as is the ability to alter the type style
(plain, italic, bold, underlined, etc.) and/or the color of tokens
in order to indicate their role in the syntax (such as which characters
are part of comments, which are part of strings, and which are other
program text).
An example is Emacs' \emph{electric font lock}.
For a token styler that handles several programming languages,
including Prolog, see \verb"m2h" by Richard O'Keefe
(\url{http://www.cs.otago.ac.nz/staffpriv/ok/software.htm}).

Nowadays, the term ``pretty-printing'' refers mainly to hardcopy
and to listings that are embedded in documentation, textbooks, and the like.
For example, consider the code in Figure~\ref{fig:sumlist}.
\begin{figure}
\hrulefill\medskip\\
\begin{verbatim}
%% sum_list(+Number_List, ?Result)
%  Unifies Result with the sum of the numbers in Number_List;
%  calls error/1 if Number_List is not a list of numbers.

sum_list(Number_List, Result) :-
    sum_list(Number_List, 0, Result).

%  sum_list(+Number_List, +Accumulator, ?Result)

sum_list([], A, A).       % At end: unify with accumulator.
sum_list([H|T], A, R) :-  % Accumulate first and recur.
  number(H),
  !,
  B is A + H,
  sum_list(Rest, B, R).
sum_list(_, _A, _R) :-    % Catch ill-formed arguments.
  error('first arg to sum_list/2 not a list of numbers').
\end{verbatim}
\hrulefill\\
\caption{Prolog code to be pretty-printed}
\label{fig:sumlist}
\end{figure}
The \LaTeX\ \verb"listings" package, which is
highly customizable and included in all \LaTeX\ distributions along with
documentation,
pretty-prints Prolog code that is surrounded with the lines:
\begin{verbatim}
\lstset{language=Prolog, frame=lines}
\begin{lstlisting}[caption={Pretty-printing example},label=pp-ex]
...
\end{lstlisting}
\end{verbatim}
The result is shown in Figure~\ref{fig:listing}.
Note how the characters of proportional-pitch type are spread apart to help
maintain horizontal alignment.

\begin{figure}
\lstset{language=Prolog, frame=lines}
\begin{lstlisting}

%% sum_list(+Number_List, ?Result)
%  Unifies Result with the sum the numbers in Number_List;
%  calls error/1 if Number_List is not a list of numbers.

sum_list(Number_List, Result) :-
    sum_list(Number_List, 0, Result).

%  sum_list(+Number_List, +Accumulator, ?Result)

sum_list([], A, A).       % At end: unify with accumulator.
sum_list([H|T], A, R) :-  % Accumulate first and recur.
  number(H),
  !,
  B is A + H,
  sum_list(Rest, B, R).
sum_list(_, _A, _R) :-    % Catch ill-formed arguments.
  error('first arg to sum_list/2 not a list of numbers').

\end{lstlisting}
\caption{Prolog listing formatted by \LaTeX\ \texttt{listings} package}
\label{fig:listing}
\end{figure}

An alternative approach to presenting Prolog programs elegantly in \LaTeX\ is
\verb"pltex" by Michael Covington (\url{http://www.ai.uga.edu/mc/pltex.zip}).
This experimental program renders the same Prolog example as shown
in Figure~\ref{fig:pltex}.
Note the use of font styles to indicate syntax, and the replacement
of \verb":-" by $\leftarrow$.  The approach is inspired by ``Algol
style'' in early computer programming literature \cite{Covington94}.
\begin{figure}
\hrulefill\medskip\\
%%%%
%
% Listing generated by PLTeX-1, M. Covington, 1997 June 15
%
\begin{flushleft}
%
% Definitions used by pltex:
%
% A box in math mode, italic type, specified width:
\def\plbox#1#2{\makebox[#2ex][l]{$\it #1 $}}
%
% Type size and spacing:
%
\def\plskipline{\vspace*{0.6em}}
\def\plsmall{\small}
\def\plbold{\bf}
%
% Special Prolog characters:
\def\planon{\hbox{\,\rule{0.6em}{1pt}\,}}                          % _
\def\plcut{\textbf{!}}                                             % !
\def\plbraceleft{\textbf{\{}}                                      % {
\def\plbraceright{\textbf{\}}}                                     % }
\def\plemptylist{\ensuremath{[\,]}}                                % []
\def\plif{\mbox{\boldmath$\leftarrow$}}                            % :-
\def\plnot{\mbox{\boldmath$\not\vdash$}}                           % \+
\def\plifthen{\ensuremath{\rightarrow}}                            % ->
\def\plexpands{\ensuremath{\longrightarrow}}                       % -->
\def\pluniv{\textrm{=.\,.}}                                        % =..
%
%
%
%     1: %% sum_list(+Number_List, ?Result)
%
\plbox{\mbox{\plsmall\rm {\%}{\%}~sum{\_}list(+Numbers{\_}List,~?Result)}}{036.1}%
\\
%
%     2: %   Unifies Result with the sum the numbers in Number_List;
%
\plbox{\mbox{\plsmall\rm {\%}}}{004.1}%
\plbox{\mbox{\plsmall\rm Unifies~Result~with~the~sum~the~numbers~in~Numbers{\_}List;}}{057.7}%
\\
%
%     3: %   calls error/1 if Number_List is not a list of numbers.
%
\plbox{\mbox{\plsmall\rm {\%}}}{004.1}%
\plbox{\mbox{\plsmall\rm calls~error/1~if~Numbers{\_}List~is~not~a~list~of~numbers.}}{056.6}%
\\
%
%     4:
%
\plskipline
%
%     5: sum_list(Number_List, Result) :-
%
\plbox{{\plbold sum{\_}list}({Numbers{\_}List\/},~{Result\/})~\plif }{044.6}%
\\
%
%     6:     sum_list(Number_List, 0, Result).
%
\plbox{}{005.4}%
\plbox{{\plbold sum{\_}list}({Numbers{\_}List\/},~{\rm 0},~{Result\/}).}{051.3}%
\\
%
%     7:
%
\plskipline
%
%     8: %  sum_list(+Number_List, +Accumulator, ?Result)
%
\plbox{\mbox{\plsmall\rm {\%}}}{003.1}%
\plbox{\mbox{\plsmall\rm sum{\_}list(+Numbers{\_}List,~+Accumulator,~?Result)}}{047.4}%
\\
%
%     9:
%
\plskipline
%
%    10: sum_list([], A, A).       % At end: unify with accumulator.
%
\plbox{{\plbold sum{\_}list}(\plemptylist ,~{A\/},~{A\/}).}{035.1}%
\plbox{\mbox{\plsmall\rm {\%}~At~end:~unify~with~accumulator.}}{034.0}%
\\
%
%    11: sum_list([H|T], A, R) :-  % Accumulate first and recur.
%
\plbox{{\plbold sum{\_}list}([{H\/}\ensuremath{|}{T\/}],~{A\/},~{R\/})~\plif }{035.1}%
\plbox{\mbox{\plsmall\rm {\%}~Accumulate~first~and~recur.}}{029.9}%
\\
%
%    12:   number(H),
%
\plbox{}{002.7}%
\plbox{{\plbold number}({H\/}),}{016.2}%
\\
%
%    13:   !,
%
\plbox{}{002.7}%
\plbox{\plcut ,}{005.4}%
\\
%
%    14:   B is A + H,
%
\plbox{}{002.7}%
\plbox{{B\/}~{\plbold is}~{A\/}+{H\/},}{014.9}%
\\
%
%    15:   sum_list(Rest, B, R).
%
\plbox{}{002.7}%
\plbox{{\plbold sum{\_}list}({Rest\/},~{B\/},~{R\/}).}{031.1}%
\\
%
%    16: sum_list(_, _A, _R) :-    % Catch ill-formed arguments.
%
\plbox{{\plbold sum{\_}list}(\planon ,~{{\_}A\/},~{{\_}R\/})~\plif }{035.1}%
\plbox{\mbox{\plsmall\rm {\%}~Catch~ill\ensuremath{-}formed~arguments.}}{029.9}%
\\
%
%    17:   error('first arg to sum_list/2 not a list of numbers').
%
\plbox{}{002.7}%
\plbox{{\plbold error}(\mbox{\tt {\char"0D}first{\char"20}arg{\char"20}to{\char"20}sum{\_}list/2{\char"20}not{\char"20}a{\char"20}list{\char"20}of{\char"20}numbers{\char"0D}}).}{058.1}%
\\
\end{flushleft}
%
% End of generated code.
%
%%%%
\hrulefill\\
\caption{Prolog listing formatted by Covington's \texttt{pltex}}
\label{fig:pltex}
\end{figure}
Also, the aforementioned \verb"m2h" program can produce \LaTeX\ output,
in addition to HTML and other output formats.  The documentation of the \LaTeX\
\verb"listings" package gives references to other pretty-printers, several
of which are applicable to Prolog.
Finally, Logtalk \cite{Moura03} includes support for several syntax
highlighters for publishing and printing both Logtalk and Prolog code.

\bigskip
\begin{minipage}{\columnwidth}
\begin{flushright}
{\small\sf
Programmers are lulled into complacency by conventions. \\
By every once in a while, by subtly violating convention, \\
you force [the maintenance programmer] to read every line \\
of your code with a magnifying glass.\\}
--- \textsc{Roedy Green} \cite{Green96-10}
\end{flushright}
\end{minipage}

%=======================================================================
\section{Naming Conventions}
\label{sec:names}
%=======================================================================

Choosing the right names is among the most crucial activities undertaken
by any software developer, especially when the programming language
is Prolog (which is untyped, with parameters that can act as input, output
or input/output and so forth).
Detailed advice on naming can be found in \cite{LedgardT87} and
\cite[p.\ 104 ff.]{KernighanP99}.

%-----------------------------------------------------------------------
\guideline{Choose a writing style for multiple-word identifiers.}
\label{guideline:first_guideline_on_choosing_names}

The basic choice is between underscore-style (\verb"like_this")
and internal capitalization (``intercaps,'' \verb"LikeThis").
A complicating factor is that Prolog requires variables to begin
with an uppercase letter and atoms to begin with a lowercase letter.
This tempts some programmers to use ``camel case'' (\verb"likeThis"),
i.e., intercaps with the first letter lowercased, which other
programmers find particularly unaesthetic.

Recall that intercaps became popular in programming languages that
did not allow underscores within names, such as Pascal and Smalltalk.
In suppressing the visual separation between words,
one seems to be reverting to the early
Middle Ages, when handwritten words were not separated
\cite[page 65]{Thompson1893},
and thus flouting an important readability tool that is a thousand years old.

Since Prolog allows underscores, and underscores resemble spaces,
there is much to be said for using underscores everywhere a word
separator is needed, both in atoms and in variable names.
The question then remains how
to capitalize variables and atoms.  The possibilities
are to stick with one case, as in \verb"Result_So_Far"
and \verb"is_boolean_function", or to choose the most appropriate case
for each individual word in a compound identifier, as in \verb"Result_so_far"
and \verb"is_Boolean_function".
The former convention is certainly more appropriate for variable names,
with the only possible exception constituted by short suffixes a program
might consistently use to make the information flow explicit, as in
\begin{verbatim}
simplify_expressions([E_in|Es_in], [E_out|Es_out]) :-
    simplify_expression(E_in, E_out),
    simplify_expressions(Es_in, Es_out).
\end{verbatim}
An alternative practice, recommended by some, is to use intercaps for
variable names and underscores within atoms.
In any case, consistency must be ensured: knowing the words used in an
an identifier and knowing whether it is an atom or a variable should
allow any team member to know how to spell it immediately.

\recommend{Use underscores to separate words in compound identifiers,
           i.e., write \texttt{is\_well\_formed}, not \texttt{isWellFormed}.
           Prefer \texttt{Result\_So\_Far} to \texttt{Result\_so\_far}.}

%-----------------------------------------------------------------------
\guideline{Make all names pronounceable.}

A predicate named \verb"stlacie" is going to confuse anyone else
reading your program ---including yourself at a later date--- even
though it may seem, at the time of writing, to be a perfectly obvious
way to abbreviate ``sort the list and count its elements.''

If you don't enjoy typing long names, type short ones.
Call your predicate \verb"sac" or even \verb"sc" while typing in
your program, then do a global search-and-replace to change it to
\verb"sort_and_count".

%-----------------------------------------------------------------------
\guideline{Avoid using different names that are likely to be pronounced alike.}

If you use \verb"foo", do not also use \verb"fou" or \verb"fu".
Many people remember pronunciations, not spellings.  Accordingly, it must be
absolutely obvious how to spell a name when all you remember is its
pronunciation.

For the same reason, do not mix up \verb"to", \verb"two", and \verb"too".
At one time it was fashionable to abbreviate ``to'' as ``2,''
thereby saving one character.  That is how computer programs
got names such as \verb"afm2tfm" (for ``AFM-to-TFM,'' a \TeX\ utility)
and DOS's \verb"exe2bin".
However, this practice creates too much confusion.  Remembering how to
spell words \emph{correctly} is hard enough: do not ask readers of your
code to remember your creative misspellings, too.
With the exception of the widely used \emph{i18n} (for ``internationalization,''
a usage coined at DEC many years ago) and \emph{L10n} (for ``localization''),
spellings like \verb"l8tr" (or \verb"l8r"?) and \verb"w1r3d" do not
facilitate communication; they just make the reader suspect that you
are still in high school.

%-----------------------------------------------------------------------
\guideline{Within names, do not express numbers as words.}

If you have three predicates for which you have no better names,
call them \verb"pred1", \verb"pred2", and \verb"pred3"
--- not \verb"pred_one", \verb"pred_two", and \verb"pred_three".
This is yet another stratagem to make spellings more predictable
from pronunciations.

Furthermore, exported predicates should not have numeric
suffixes unless the number is some sort of code number.
For example,
\begin{verbatim}
unicode_4_0_0(?Code, ?Class)
\end{verbatim}
might be reasonable.

%-----------------------------------------------------------------------
\guideline{Choose sensible names for auxiliary predicates.}

If part of the algorithm for your predicate, which you have named
\verb"foo", needs to be placed in another predicate definition,
do not immediately call that predicate \verb"foo_aux";
usually there are better alternatives.
For example:
\begin{itemize}
\item
If the auxiliary predicate and the main predicate have different
numbers of arguments, their names can be the same.
\item
If an auxiliary predicate is there to do a case analysis,
you can use \verb"_case" as a better suffix than \verb+_aux+.
\item
If an auxiliary predicate is a loop, \verb+_loop+ may be a good suffix.
\item
If an auxiliary predicate relies on a higher predicate
to take care of resource allocation and disposal,
\verb+_unguarded+ might be a good suffix, as in:
\begin{verbatim}
foo(...) :-
    acquire_resources(...),
    call_cleanup(foo_unguarded(...), release_resources(...)).

foo_unguarded(...) :-
    ...
\end{verbatim}
\end{itemize}
When all else seems inappropriate \verb"_aux" can be used.
Other arbitrary alternatives are \verb"_1", \verb"_2", \verb"_3" and so on
or \verb"_x", \verb"_xx" and so forth, but these give up any opportunity to
indicate the function of the auxiliary predicate and so should
preferably be accompanied by explanatory comments.

%-----------------------------------------------------------------------
\guideline{If a predicate represents a property or relation, its name
  should be a noun, noun phrase, adjective, prepositional phrase, or
  indicative verb phrase.}

Examples include:
\begin{itemize}
\item
\verb"sorted_list", \verb"well_formed_tree", \verb"parent"
(nouns or noun phrases);
\item
\verb"well_formed", \verb"ascending" (adjectives);
\item
\verb"in_tree", \verb"between_limits" (prepositional phrases);
\item
\verb"contains_duplicates", \verb"has_sublists" (indicative verb phrases).
\end{itemize}

%-----------------------------------------------------------------------
\guideline{If a predicate is understood procedurally ---that is, its
  job is to do something, rather than to verify a property--- its name
  should be an imperative verb phrase.}

Examples include
\verb"remove_duplicates" (i.e., the second-person imperative form
is used, not \verb"removes_duplicates") and
\verb"print_contents" (not \verb"prints_contents").

%-----------------------------------------------------------------------
\guideline{Choose predicate names to help show the argument order.}
\label{guideline:mother_child}

For example, \verb"mother_of(A,B)" is ambiguous;
does it mean ``A is the mother of B'' or ``the mother of A is B''?
Naming it \verb"mother_child"
or \verb"mother_of_child" would eliminate the ambiguity.

In general, the predicate name is the only place in Prolog where you can
conveniently give the user hints about the argument order.
For example, a specification like \verb"tree_size(+T, ?S)"
makes it clear that the tree
argument precedes the size argument.  This means that sometimes the
most salient type name needs to come last.  For example, to convert
lists to trees, we might use \verb"list_to_tree(+L, ?T)".
Module prefixes, on the other hand, can only be prefixes.
If you see a goal of the form \verb"tree:from_list(X, Y)",
can you tell which of the arguments is the tree and which the list?
Hence, \verb"tree:tree_from_list(X, Y)" should not be seen
as gratuitously redundant.

%-----------------------------------------------------------------------
\guideline{Use descriptive names for variables wherever possible, and
  make them accurate.}

For example, do not call a variable \verb"Tree" if it is not a
tree. You would be surprised at how often such things are done by
programmers who have changed their minds midway through constructing a
clause.

More generally, there are two things that can be encoded
in the variable name: its type and its role in the predicate.
You can name the variable after only one of these properties
if the other is obvious from the context.  For example,
if you have a predicate \verb"graph_nodes(+Graph, -Nodes)",
it is already obvious that the role of the second argument
is to represent graph nodes, so you may name it \verb"List",
if it is a list, or \verb"Assoc", if it is an association list;
for maximum clarity you may even name it \verb"Node_List"
or \verb"Node_Assoc".

%-----------------------------------------------------------------------
\guideline{Use single letter names only in presence of suitable conventions.}

Prolog has no global variables; most variables exist only in a very limited
context.
For conciseness, some programmers adopt program-wide conventions whereby,
for instance, \verb"N" is always
the total number of elements to process and \verb"I" is always the number
of elements processed so far.
If you have a convention that
\verb"L" and \verb"U" are always the lower and upper bounds of some range,
fine.
If you have a convention that \verb"C", \verb"C0", \verb"C1", \verb"C2"
are always character codes, \verb"S", \verb"S0", \verb"S1", \verb"S2",
are sequences, and so on, fine;
make sure that such conventions are prominently stated in the source file
documentation or are absolutely obvious from context.
Whenever you have the slightest doubt that the code may be tricky to read,
forget about single letter names and give your reader some help.

%-----------------------------------------------------------------------
\guideline{Consistently name threaded state variables.}

There is a special case for variables that receive an initial state,
go through some translations, and come to a final output.
It is important to name those consistently.
For example,
one can use names of the forms \verb"State0" (for the initial state),
\verb"State1", \verb"State2" and so on (for the intermediate states),
and \verb"State" (for the final state):
\begin{verbatim}
foo(..., State0, State) :-
    foo_step(State0, State1),
    foo(..., State1, State).
\end{verbatim}
An alternative choice is to use the names
\verb"State_in" (for the initial state),
\verb"State_tmp", \verb"State_tmp1", \verb"State_tmp2" and so forth
(for the intermediate states),
and \verb"State_out" (for the final state):
\begin{verbatim}
foo(..., State_in, State_out) :-
    foo_step(State_in, State_tmp),
    foo(..., State_tmp, State_out).
\end{verbatim}

%-----------------------------------------------------------------------
\guideline{Use a singular noun for the first element of a list and its
  plural for the remaining elements.}
\label{guideline:last_guideline_on_choosing_names}

For example, match a list of trees to \verb"[Tree|Trees]".
If you use a single letter for a list element,
use that letter with an `s' after it for the remaining elements.
For example, match a list of trees to \verb"[T|Ts]".
Only do this if something nearby in the context, such as
an adequately detailed comment, makes it obvious
what the single letter stands for.
A compromise, which may be more readable in some situations,
is \verb"[T|Trees]".

If there are two entities
for which you would like to use the same single letter, give
them both full names, as in
\begin{verbatim}
term_translations([], []).
term_translations([Term|Terms], [Translation|Translations]) :-
    term_translation(Term, Translation),
    term_translations(Terms, Translations).
\end{verbatim}
Here, using \verb"T" would be confusing.

%=======================================================================
\guideline{Decide whether predicate names should carry the types on which they operate}

The reason a decision is necessary is that, when you also have module names,
you can end up with redundancies such as
\verb"quaternion:quaternion_magnitude(Q, A)".
If you are using a module system, and you are always including the module name,
then you need not put the same information in the predicate names.

On the other hand,
a number of widely accepted libraries use predicate names that
carry the types on which they operate.  This is the case, e.g.,
for the ordered set library (using the \verb"ord_" prefix)
and the association lists library (using the \verb"_assoc" suffix).
This convention predates the module system and is similar to a
long-standing convention in C, also predating other methods of managing the
namespace.

Even when modules are used, some people like to import predicates from
other modules and use them without module prefixes, just like you use
predicates provided by the system.  This is the historically preferred
approach, because people expected to convert Edinburgh Prolog code that
used plain files to code that used modules just by adding
\verb":- use_module" directives.  In this style, type names are still
useful even if they are identical to
the module name.

%=======================================================================
\guideline{Do not blindly import styles from other languages.}

For instance, do not blindly follow the Java tradition by calling everything
in sight \verb"is_xxx" or \verb"get_yyy".
It is fine to name \verb"is_tree/1" a \emph{checker}, that is, a predicate
that will only succeed, without leaving choice-points, on well-formed,
structurally complete trees.  This will help the user to distinguish
it from the \emph{generator} ---whether its existence in the program
is real or only conceivable--- called \verb"tree/1".

Similarly, stay away from the massive overuse of ``get'' that stems from
the Java world.  It is fine to use ``get'' to signal that information is being
obtained
in an extra-logical way, such as when I/O or the foreign
language interface are involved.
But do not use ``get'' for fetching a property via an ordinary
logical relation.

\bigskip
\begin{minipage}{\columnwidth}
\begin{flushright}
{\small\sf
Much of the skill in writing unmaintainable code \\
is the art of naming variables and methods. \\
They don't matter at all to the compiler. \\
That gives you huge latitude to use them \\
to befuddle the maintenance programmer.\\}
--- \textsc{Roedy Green} \cite{Green96-10}
\end{flushright}
\end{minipage}

%=======================================================================
\section{Documentation}
\label{sec:documentation}
%=======================================================================

All the good reasons why documentation is crucial to the success of
any software project, independently from the languages used to code it,
are valid for Prolog code.  In addition,
features like multiple modes, dynamic code, meta-programming
facilities, and the lack of prescriptive typing
make documentation even more important than for other
programming languages.

On the other hand, writing good documentation is expensive and this
cost may not be fully justified for those projects where Prolog is used for
rapid prototyping.  Each project should choose a consistent
documentation standard based on the expected benefit/cost ratio.
In this choice an important role can be played by the available
tools: if the code is written with a system that provides good support
for writing, checking, formatting and retrieving the documentation,
it is quite likely that choosing the documentation standard(s) dictated
by that system will be the most economical solution.

In this section, we will review the important points to consider when
designing a code documentation standard.

%-----------------------------------------------------------------------
\guideline{Begin every predicate (except perhaps auxiliary predicates) with an introductory comment in a well-defined format.}

Predicates that can be called from elsewhere in the program
---from code written by others, or by yourself on a different occasion---
must be documented by an introductory comment.
Here is an example:
\begin{verbatim}
%% remove_duplicates(+List, -Processed_List) is det
%
%  Removes the duplicates in List, giving Processed_List.
%  Elements are considered to match if they can
%  be unified with each other; thus, a partly uninstantiated
%  element may become further instantiated during testing.
%  If several elements match, the last of them is preserved.
\end{verbatim}

\begin{figure}
\hrulefill\medskip\\
\begin{verbatim}
%% <name>(<mode><variable>[:<type>], ...) is <determinism>
%
%  <description line 1>
%  <description line 2>
%  ...

<name>(<variable>, ...) :-
    ...
\end{verbatim}
\hrulefill\\
\caption{Example documentation template}
\label{fig:documentation-template}
\end{figure}
This comment follows the template in
Figure~\ref{fig:documentation-template}.
Note the overall layout and the use of \verb"%%" to mark the first line.
Arguments are represented by variable names preceded by mode specifiers
from one of the systems in Table~\ref{tab:mode-specifiers},
and optionally followed by type specifiers.
At the end of the first line is a determinism specifier
from Table~\ref{tab:determinism-specifiers}.

\begin{table}[h!]
\caption{Three systems of argument mode specifiers}\begin{flushleft}
\hrulefill\medskip\\
Recommended system:\\
  \begin{tabular}{ l p{10cm} }
    \verb"*" & \verb"ground" on entry, i.e., must already
    	       be instantiated to a term that contains no uninstantiated
               variables when the predicate is called.
               Thought of as \emph{input}. \\
    \verb"+" & \verb"nonvar" on entry, i.e., must already be instantiated
               to a term that is not an uninstantiated variable (although
               the term may contain an uninstantiated variable) when the
               predicate is called.
               Thought of as \emph{input}. \\
    \verb"=" & Thought of as \emph{input} but may be \verb"var" on entry,
               i.e., may or may not be instantiated when the predicate is called.
               This argument should not share variables with any
               argument that might be further instantiated by this predicate;
               this one will be left as it was found. \\
    \verb"-" & \verb"var" on entry, i.e., must be uninstantiated when
               the predicate is called.
               Thought of as \emph{output}. \\
    \verb"/" & Similar to `\verb"-"': this argument should be an unbound
                variable that, in addition, is not shared with any other
                argument. \\
    \verb">" & Thought of as \emph{output} but might be \verb"nonvar" on entry.
    	           Instantiation must not control the behavior of the predicate. \\
    \verb"?" & Not specified, i.e., may or may not be instantiated when the
               predicate is called.
    	           Instantiation may control the behavior of the predicate.
               Thought of as either \emph{input} or \emph{output}
               or \emph{both} input and output. \\
  \end{tabular}\\
~\\
System used in PlDoc (see below): \\
  \begin{tabular}{ l p{10cm} }
    \verb"+" & \verb"nonvar" on entry, i.e., must already be instantiated
               to a term that is not an uninstantiated variable (although
               the term may contain an uninstantiated variable) when the
               predicate is called.
               Thought of as \emph{input}. \\
    \verb"-" & \verb"var" on entry, i.e., must be uninstantiated when
               the predicate is called.
               Thought of as \emph{output}. \\
    \verb"?" & Not specified, i.e., may or may not be instantiated upon entry.
               Thought of as either \emph{input} or \emph{output}
               or \emph{both} input and output. \\
    \verb":" & A meta-argument. Implies \verb"+". \\
    \verb"@" & Not further instantiated. Typically used for type-tests. \\
    \verb"!" & Contains a mutable structure that may
               be modified using \verb"setarg/3"
               or \verb"nb_setarg/3". \\
  \end{tabular}\\
~\\
Simplest system:\\
  \begin{tabular}{ l p{10cm} }
    \verb"+" & \verb"nonvar" on entry (normally), i.e., will normally be
               instantiated to a term that is not an uninstantiated variable
               (although the term may contain an uninstantiated variable)
               when the predicate is called.
               Thought of as \emph{input}. \\
    \verb"-" & \verb"var" on entry (normally), i.e., will normally be
               uninstantiated when the predicate is called.
               Thought of as \emph{output}. \\
    \verb"?" & Not specified, i.e., may or may not be instantiated when the
               predicate is called.
               Thought of as either \emph{input} or \emph{output}
               or \emph{both} input and output. \\
\end{tabular}\medskip\\
\hrulefill\\
\end{flushleft}
\label{tab:mode-specifiers}
\end{table}

\paragraph{Argument mode specifiers.}
The first of the three systems in
Table~\ref{tab:mode-specifiers} is preferred
because it is the most explicit about flow of information.
For example, note that it would be wrong to write
\begin{verbatim}
compare(?R:order, +T1:term, +T2:term)
\end{verbatim}
to document the \verb"compare/3" primitive,
because \verb"T1" and \verb"T2" are allowed to be variables, and
\begin{verbatim}
compare(?R:order, ?T1:term, ?T2:term)
\end{verbatim}
would be misleading because we expect that \verb"?" arguments will
be unified with something.
An annotation in this scheme that captures the sensible uses of
\verb "compare/3" is
\begin{verbatim}
compare(?R:order, =T1:term, =T2:term)
\end{verbatim}

% The difference between \verb">" and \verb"?" is that a \verb"?"
% argument may be allowed to control what the predicate does, but a
% \verb">" argument should not.

\paragraph{Argument type specifiers.}
The type specifiers, if included, can be arbitrary types meaningful to the
reader. Informal types such as
\verb"atomic", \verb"integer", \verb"list" and \verb"string" are fine,
but it is recommended that any unconventional interpretations be documented
near the start of the file, program or application as appropriate.
Completely informal descriptions can only work
for simple and/or small projects.  In more complex cases a BNF-like
notation can be adopted or, better, one can borrow the syntax of some
formal type system, such as the ones described in
\cite{FagesC01,Hermenegildo00,MycroftO84,JefferyHS00}.\footnote{It has not
escaped our notice that the use
of formal types in code documentation immediately suggests the
possibility of giving the compiler access to them in order to
enable useful diagnostics of all kinds, if not optimized compilation.
Even though we believe that adding standard ways of \emph{optionally}
specifying prescriptive types
and other annotations (e.g., modes, determinism, cost, termination)
to Prolog would result in a
significantly more powerful language, all this is completely beyond
the scope of this paper.}

\begin{table}
  \caption{Predicate determinism specifiers for the template
           of \textup{Figure~\ref{fig:documentation-template}}}
  \label{tab:determinism-specifiers}
\begin{flushleft}
\hrulefill\medskip\\
  \begin{tabular}{ l p{9cm} }
    \verb"det"     &   Must succeed exactly once and leave
                       no choice-points. \\
    \verb"semidet" &   Must either fail or succeed exactly once and leave
                        no choice-points. \\
    \verb"multi"   &   Must succeed at least once but may leave choice-points
                        on the last success. \\
    \verb"nondet"  &   May either fail or succeed any number of times and
                        may leave choice-points on last success.
  \end{tabular}\medskip\\
\hrulefill\\
\end{flushleft}
\end{table}

\paragraph{Predicate determinism specifiers.}
The degree of determinism of a Prolog predicate
is an important part of its behavior and so must always be
documented.  Doing so forces the author to consciously consider the
determinism and ensures that this information is available for other
users of the predicate.  Furthermore, by specifying determinism in a
standardized representation, like the one defined in
Table~\ref{tab:determinism-specifiers},
this information is made available to
automated test tools.  Informally, \verb"det" is used for deterministic
transformations (e.g., arithmetic), \verb"semidet" for tests,
\verb"nondet" and \verb"multi" for generators.

\vfill

After the predicate specification comes a clear explanation, in natural
language, of what the predicate does, including special cases (in the
\verb"remove_duplicates/2" example, lists with uninstantiated elements).
Normally, you should write the comment before constructing the
predicate definition; if you cannot describe a predicate coherently
in natural language, then you are not ready to write it.

The basic template can easily be extended to deal with multiple modes,
e.g.,
\begin{verbatim}
%% age(+Name:atom, -Age:integer) is semidet
%% age(-Name:atom, +Age:integer) is nondet
%
%   ...
\end{verbatim}
Similarly, it is often convenient to document tightly related predicates
with multiple arities using just one documentation block, as in
\begin{verbatim}
%% rdf_load(+File)
%% rdf_load(+File, +Options)
%
%  ...

rdf_load(File) :-
    rdf_load(File, []).

rdf_load(File, Options) :-
    ...
\end{verbatim}

When a documentation system supported by adequate
tools is available, it should be given appropriate consideration.
For instance, if you use SWI-Prolog, it may be advantageous to use
PlDoc, the SWI-Prolog source-code documentation infrastructure
(\url{http://www.swi-prolog.org/pldoc/package/pldoc.html}).
The comment template used by PlDoc is not very different from the one
exemplified in this guideline.
It uses the second set of argument mode specifiers, and it
also provides tags such as \verb"@param", \verb"@throws", and \verb"@error"
for further description of a predicate's requirements and behavior.

%-----------------------------------------------------------------------
\guideline{Use comments to make main predicates visibly distinct
from auxiliary predicates.}

\emph{Auxiliary predicates} exist only to continue the definition of
another predicate, and are not called from anywhere else.  For
example, recursive loops and subordinate decision-making procedures
are often placed in auxiliary predicates. While users of your code need not
know about auxiliary predicates, commenting predicates can
provide useful technical documentation for maintenance and reuse.

Begin the first line of comments with \verb"%%" when introducing
a predicate called from elsewhere in the program, but \verb"%" when
introducing an auxiliary predicate.
For example, we can have
\begin{verbatim}
%% remove_duplicates(+List, -Processed_List) is det
%
%  Removes the duplicates in List, giving Processed_List.
%  ...
\end{verbatim}
and
\begin{verbatim}
%   remove_duplicates_loop(+List, -Processed_List) is det
%
%   Processed_List contains the first occurrence of each element
%   of List, in the same order.
\end{verbatim}

Auxiliary predicates in Prolog often correspond to loops in other
languages; the meaning of the Prolog predicate is the loop invariant,
and really should be stated in a comment.

%-----------------------------------------------------------------------
\guideline{Use descriptive argument names in the introductory comment.}

The argument names \verb"Index", \verb"List" and \verb"Elem" in the
introductory comment of the predicate \verb"nth0/3" below were chosen
to convey its intended usage.
\begin{verbatim}
%% nth0(?Index, ?List, ?Elem)
%
%  True if Elem is the Index'th element of List.
%  Counting starts at 0.

nth0(Index, List, Elem) :-
    ...
\end{verbatim}

Well-chosen argument names can also hint at the expected type of
arguments, which is particularly important if explicit argument types
are not specified. For example, it is reasonable to expect the
argument \verb"List" to be associated with some representation of a
list and, in the absence of any other information, this would most
likely be the usual Prolog list representation. Likewise,
\verb"Index" would, in the absence of any other information, suggest a
numeric argument type.

%-----------------------------------------------------------------------
\guideline{Argument names in the clauses should preferably be
           the same as in the introductory comment.}

Where practical, use the same argument names in the clause definitions
as used in the introductory comment, like this:
\begin{verbatim}
%% nth0(?Index, ?List, ?Elem)
%
%  True if Elem is the Index'th element of List.
%  Counting starts at 0.

nth0(Index, List, Elem) :-
    ...
\end{verbatim}
However, there will be frequent occasions when this is not practical;
in the following example, \verb"List" in the comment corresponds
to \verb"[X|_]" in one clause and \verb"[_|T]" in the other:
\begin{verbatim}
%% member(?Elem, ?List)
%
%  True if Elem is a member of List.

member(X, [X|_]).
member(X, [_|T]) :-
    member(X, T).
\end{verbatim}

%-----------------------------------------------------------------------
\guideline{If code needs elaborate explanation, consider rewriting it.}

Prolog lends itself to simple, logical programs:
\emph{one predicate, one idea.}
The act of producing documentation is helpful also because it gives
you a chance to spot design mistakes:
if you use `and' several times while describing a single predicate,
that predicate is doing too many jobs and should be broken up.

\bigskip
\begin{minipage}{\columnwidth}
\begin{flushright}
{\small\sf
You don't have to actively lie,\\
just fail to keep comments as up to date with the code.\\}
--- \textsc{Roedy Green} \cite{Green96-10}
\end{flushright}
\end{minipage}

%=======================================================================
\section{Language Idioms}
\label{sec:language-idioms}
%=======================================================================

These guidelines are particularly addressed to those who are
relatively new to Prolog and accustomed to the practices of other
programming languages.  In particular, we highlight the importance of
\emph{steadfastness} and tail recursion, the dangers of abusing cuts
and the database, the choice of the right kind of lists depending on the
operations to be efficiently supported,

%-----------------------------------------------------------------------
\guideline{Predicates must be steadfast.}

Any decent predicate must be ``steadfast,'' i.e., must work correctly
if its output variable already happens to be instantiated to the
output value \cite{OKeefe90}.
That is,
\begin{verbatim}
?- foo(X), X = x.
\end{verbatim}
and
\begin{verbatim}
?- foo(x).
\end{verbatim}
must succeed under exactly the same conditions and have the same side effects.
Failure to do so is only tolerable for auxiliary predicates whose call
patterns are strongly constrained by the main predicates.

%-----------------------------------------------------------------------
\guideline{Place arguments in the following order: inputs, intermediate results, and final results.}
\label{guideline:ordering_of_arguments}

In any programming language, it can get confusing when a procedure
has more than a few parameters.  C's \verb"stdio"\footnote{Part of the C
standard library providing various standard input and output operations.}
is famously confusing
because it sometimes puts the stream argument first and sometimes
last.  Programming language designers know of three things that
can help:
\begin{enumerate}
\item
Strong types help the compiler notice when parameters are in
the wrong order.  Prolog does not have such a type system, although
the related languages G\"odel \cite{HillL94}
and Mercury \cite{SomogyiHC95} do, and the DEC-10 Prolog type
checker \cite{MycroftO84} has inspired several
type systems for Prolog.
However, even a type system does not help when two arguments have the
same type.  What would have been a single ``reference parameter''
in \Cplusplus{} or ``in out'' parameter in Ada or Fortran is a
before/after pair of parameters in Prolog.
\item
Keywords are used in many languages, including Common Lisp, Ada,
S, DEC Pascal, Mesa, SML (in a sense), and Smalltalk.
As yet no keyword facility
for Prolog has become well known, let alone widely accepted.
Some people have used ``wrappers'', writing code like
\begin{verbatim}
    generalised_hanoi(pegs(5), discs(10), solution(S))
\end{verbatim}
Single-constructor single-argument types in SML and Haskell can be used
this way, but those compilers have been let into the secret.  Prolog systems
take these wrappers literally and build run-time structures for them, which
hurts performance.  The big problem is indexing, which usually does not look
inside wrappers.  This style cannot be recommended.
\item A consistent convention is the only alternative left.
\end{enumerate}

You can of course try to devise your own convention.  The one described
here is a simplification of the one presented in \cite[pp.\ 12--16]{OKeefe90},
which was constructed to be consistent with the built in predicates of
then-current Edinburgh Prologs.

To be useful, an argument order convention must not only be consistent,
it must be \emph{memorable}, which suggests that it should be built on
some sort of natural metaphor.  The one used here is Space-Is-Time.
If $X$ is known or needed before $Y$ in time, it is placed before $Y$
in the arguments.  If $X$ represents an earlier state and $Y$ represents a
later state, then $X$ is earlier in the arguments.

The basic rule then is ``inputs before outputs,'' which is compatible
with first-argument indexing.  When the computer is
choosing a clause, or a human is reading a predicate, some of the
arguments are more important than others, typically the argument(s)
controlling a recursion.  These being needed before the others (which
might be context parameters or accumulators) are written before them.

Let us consider a little example.  We want to walk over a binary tree,
collecting labels that include a given flag.  There will be four
parameters.  We will be using a \verb"Labels0"\dots\verb"Labels"
pair to build the
list.  There will be the \verb"Flag" we want to look for.  And there will
be the \verb"Tree".  The \verb"Labels" are an output, so that pair go last.
Within that pair, \verb"Labels0" precedes \verb"Labels" in the list,
so it precedes \verb"Labels" in the parameters.
The \verb"Tree" drives the induction, so it is the input we need first.
Hence there is no choice here:
\begin{verbatim}
items_including(Tree, Flag, Labels) :-
    items_including(Tree, Flag, Labels, []).

items_including(empty, _, Labels, Labels).
items_including(node(Label,Left,Right), Flag, Labels0, Labels) :-
    items_including(Left, Labels0, Labels1),
    (member(Flag, Label) ->
        Labels1 = [Label|Labels2]
    ;
        Labels1 = Labels2
    ),
    items_including(Right, Labels2, Labels).
\end{verbatim}

There is a noteworthy and bitterly regrettable exception to the
``most discriminating inputs first'' rule, and that is input/output
commands with a Stream argument.  These commands were introduced
at Quintus without any internal discussion.  A vastly better
design would have been something like MIT Scheme's
\verb"with-input-from-port" and \verb"with-output-from-port", making it easy
for programmers to write composite input/output commands that could
be redirected easily without \emph{any} stream argument threading.
It would not only have been easier to use, it would have been
more efficient, because Stream arguments have to be checked every
time they are used.  Indeed, a preliminary input redirection
facility of this kind was in the Quintus library when
\verb"write(Stream, Term)" was introduced.  Why do we mention this here?
Because there are two important style lessons:
\begin{enumerate}
\item
Get ideas from more than one other language.  Stream arguments
were a direct copy from C.  The better alternative is found in Scheme
and Ada.  Scheme and functional languages are often good sources of
ideas for Prolog; C and Java are less likely to have ideas that fit
well into a (mostly-)declarative framework.
\item
Talk to other people \emph{before} you release.
\end{enumerate}

This raises the question of optional arguments.  Sometimes a family
of predicates can be thought of as a single predicate with optional
arguments.  There are two patterns you can use:
\begin{itemize}
\item required inputs, optional inputs, outputs;
\item inputs, required outputs, optional outputs.
\end{itemize}
Sometimes you might technically be able to have both optional
inputs and optional outputs, relying on some required input to
discriminate, but people reading the code will have a hard time.
Don't do that.  Stream arguments are again an exception to the
rule, but the Options in \verb|open| and \verb|write_term| and
so on show the norm.

The other major exception to the inputs before outputs pattern
is \verb|is/2|.  This follows the convention of assignment commands
in most programming languages, so should not be hard to remember.
Note, however, that when SWI-Prolog lets you call your own predicates
in arithmetic expressions, it requires them to report their result
via their \emph{last} argument, conforming to the usual pattern.

%-----------------------------------------------------------------------
\guideline{Try to limit the number of predicate arguments.}
\label{guideline:try_to_limit_the_number_of_predicate_argument}

A predicate with too many arguments may be trying to do several
conceptually separate jobs at once,
or some of the arguments may need to be replaced with data structures
containing pieces of information bundled together.

%-----------------------------------------------------------------------
\guideline{Use cuts sparingly but precisely.}

First think through how to do the computation \emph{without} a cut;
then add cuts to save work.
For further guidance see \cite[pp.\ 88--101]{OKeefe90}.
Concerning code layout, make sure cuts do not go unnoticed:
if a green cut\footnote{A \emph{green cut} is one that only prunes away
some computation paths not contributing new solutions.
A \emph{red cut} is a cut that is not green.}
may be placed on the same line as the previous predicate call,
red cuts definitely must be on their own line of code.

%-----------------------------------------------------------------------
\guideline{Never add a cut to correct an unknown problem.}

A common type of Prolog programming error is manifested in a predicate
that yields the right result on the first try but goes wrong upon
backtracking.  Rather than add a cut to eliminate the backtracking,
investigate what went wrong with the logic.  There is a real risk that
if the problem is cured by adding the cut, the cut will be far away
from the actual error (even in a different predicate), which will
remain present to cause other problems later.

%-----------------------------------------------------------------------
\guideline{Work at the beginning of the list.}

You can get to the first element of a 1000-element list in one step;
getting to the last element requires 1000 steps.  But do not let this
or any other guideline stop you from doing the computation that you
need to do.  In some cases you can speed up a computation by building
a list in the reverse of the order you first thought of.

%-----------------------------------------------------------------------
\guideline{Avoid {\tt append} where speed is important.}

Generally, \verb"append" is slow because it has to go all the way to the end of the first list before adding a pointer to the second.
Use it sparingly in speed-critical code.
But don't be silly: do not re-implement \verb"append" with another name
just to avoid using the original one.

One good tactic is to use \verb"append" liberally in prototyping, and
then change to other methods, such as difference pairs, in the
finished implementation.

%-----------------------------------------------------------------------
\guideline{Use difference pairs (difference lists) to achieve fast concatenation.}

Suppose you want to combine the sequences $\langle a,b,c \rangle$ and
$\langle d,e,f \rangle$ to get $\langle a,b,c,d,e,f \rangle$. One way
is to append the lists \verb"[a,b,c]" and \verb"[d,e,f]".

But there's another way. Store the sequences as lists with
uninstantiated tails, \verb"[a,b,c|X]" and \verb"[d,e,f|Y]", and keep
copies of \verb"X" and \verb"Y" outside the lists (so you can get to
them without stepping through the lists, which is what \verb"append"
wastes time doing).
Then just instantiate \verb"X" to your second list,
\verb"[d,e,f|Y]". {\it Voil\`{a}\/}: the first list is now
\verb"[a,b,c,d,e,f|Y]"; you have added material at the end by
instantiating its tail. Later, by instantiating \verb"Y" to something,
you can add even more elements.  The applicability of this technique
is limited by the fact that difference pairs can only be appended once.

Difference pairs are discussed in, e.g., \cite{CovingtonNV97}
and \cite{OKeefe90}.

%-----------------------------------------------------------------------
\guideline{Recognize the memory advantages of tail recursion.}

Tail recursion is recursion in which the recursive call is the last
subgoal of the last clause, leaving no backtrack points.  In this
circumstance, there is no need to create a new environment frame on
the stack: control can be directly transferred to the beginning of the
predicate.
See, e.g., \cite[Chapter~4]{CovingtonNV97}.

The memory saved by tail recursion can be very important when a
recursive loop has to go through thousands or millions of cycles, such as
processing single characters of a large input file.
When the recursion depth is measured in dozens or less, the memory saved by
tail recursion is not important.  Moreover, not all computations can easily
be made tail recursive.
Accordingly, do not be afraid of body recursion (non-tail recursion)
if you need it.

Ordinarily, in your first version of a program, you should use tail
recursion if you can easily do so, but use body recursion if that
makes it easier to get the code right.  Then change body recursion to
tail recursion if you find that there is a substantial advantage to
doing so.

%-----------------------------------------------------------------------
\guideline{Avoid {\tt asserta}/{\tt assertz} and {\tt retract} unless you actually need to preserve information through backtracking.}

Although it depends on your compiler, \verb"asserta"/\verb"assertz"
and \verb"retract"
are usually very slow.  Their purpose is to store information that
must survive backtracking. If you are merely passing intermediate
results from one step of a computation to the next, use arguments.

If you have a dynamic predicate, write interface predicates for
changing it instead of using 'bare' calls to \verb"asserta"/\verb"assertz"
and
\verb"retract", so that your interface predicates can check that the
changes are logically correct, maintain mutexes for multiple threads,
and so forth.

%-----------------------------------------------------------------------
\guideline{Consider doing things in batches rather than one at a time.}

This is language-independent advice.  The reason for it is that
if you have several things to do to one data structure, you can
often share the overheads among a group of operations.

Balanced binary trees typically take about twice as
much memory as lists, but they are superb for inserting,
deleting, and testing for the presence of single elements in
$O(\log n)$ worst case time.  You can compute the union,
intersection, set difference, and so on of sets of size
$m$ and $n$ in $O(m \log n)$ time.  But
using sorted lists, you can do these operations in $O(m+n)$ time
(see \cite{OKeefe90}).
This is why \verb"setof/3" returns a sorted list of answers.

The ``merge'' paradigm can be used any time you have collections
sorted on some key to be combined in some way.  For example, if
you represent polynomials as lists of Exponent-Coefficient pairs,
the usual algorithm for adding them is a merge.

This makes sorting an important efficiency tool in Prolog.
Beware of comparing non-ground terms, however.

%-----------------------------------------------------------------------
\guideline{For sorting, use merge sort or a built-in sorting algorithm.}

Several Prolog books show how ``cute'' quicksort is in Prolog,
using it as an example of the use of list differences, but fail
to point out that merge sort is better for Prolog (and for other languages,
such as ML and Haskell).

Merge sort, an $O(n\log n)$ sorting algorithm, generally outperforms
quicksort, which is also $O(n\log n)$ in the average case but $O(n^2)$
in a surprisingly common worst case.  A Prolog implementation of
merge sort is given in \cite{CovingtonNV97}. 
A bottom-up merge sort that is $O(n)$ for
already sorted input is given in \cite{Brady05}.

Quicksort was invented for an Elliot~405 computer
with 512 words of memory \cite{Hoare62}.  The performance comparison in that
paper is of a drum-based quicksort with a drum-based merge sort.
The chief virtues of quicksort are that it does
not require extra workspace and that
the inner loop is fast \emph{if} comparisons
are exceedingly cheap.  Neither advantage exists when sorting
linked lists.  As Hoare showed, quicksort does significantly
more comparisons than merge sort.

Perhaps more interestingly, merge sort gives you the chance to
combine the values of key-value pairs on the fly; because
quicksort often puts records with equal keys in different
partitions, quicksort variants have to wait until the end.

It should be stressed that many Prolog systems have efficient built-in
sorting algorithms that are likely to be faster than anything you
would have written, if they fit your needs.

%-----------------------------------------------------------------------
\guideline{Develop your own \emph{ad hoc} run-time type and mode checking system.}

Many problems during development (especially if the program is large
and/or there are several developers involved) are caused by passing
incorrect arguments.
Even if the documentation is there to explain, for each predicate,
which arguments are expected on entry and on successful exit, they
can be, and all too often they are, overlooked or ignored.
Moreover, when a ``wrong'' argument is passed, erratic behavior can manifest
itself far from where the mistake was made (and of course, following
Murphy's laws, at the most inconvenient time).

In order to significantly mitigate such problems, do take the
time to write your own predicates for checking the legality of arguments
on entry to and on exit from your procedures.
In the production version, the goals you added for these checks
can be compiled away using \verb"goal_expansion/2".

\bigskip
\begin{minipage}{\columnwidth}
\begin{flushright}
{\small\sf
Global variables save you from having to specify arguments in \\{}
[predicates]. Take full advantage of this. Elect one or more of these \\
global variables to specify what kinds of processes to do on the \\
others. Maintenance programmers foolishly assume that [Prolog \\
predicates] will not have side effects.\\}
--- \textsc{Roedy Green} \cite{Green96-10}
\end{flushright}
\end{minipage}

%=======================================================================
\section{Development, Debugging, Testing}
\label{sec:development}
%=======================================================================

The fundamental principles of reliable program development and testing
are the same for all programming languages, but the details of how to
apply these principles in Prolog involve ideas that will be new to
those who have only worked with conventional languages.  In
particular, unification, backtracking, and the concept of success
vs.\ failure all introduce new twists to the art of program debugging.

%-----------------------------------------------------------------------
\guideline{Invest appropriate (not excessive) effort in the program; distinguish a prototype from a finished product.}

Many Prolog programs are exploratory; when trying to figure out whether
something can be done, it is reasonable to
do each part of it in the easiest possible way, whether or not it is efficient.

%-----------------------------------------------------------------------
\guideline{The most efficient program is the one that \emph{does the right computation,} not the one with the most tricks.}

Using efficiency tricks, you can sometimes double the speed of a
computation. By choosing a better basic algorithm, you can sometimes
speed a computation up by a factor of a million.

Do not modify your code just for the sake of efficiency until you are
\emph{sure} it is actually doing the right computation.  Of course
there may be more than one ``right'' computation, so keep in mind that
simplicity often leads to speed.

%-----------------------------------------------------------------------
\guideline{When efficiency is critical, make tests.}

Don't believe what people tell you (including us); do your own experiments.
The built-in Prolog predicate \verb"statistics" will tell you the time and memory used by a computation.
Many Prolog compilers provide other predicates for measuring efficiency. If available in your Prolog environment,
use the profiler to measure and analyze efficiency.

Use the debugger, especially a graphical one if available, to examine
choice points (backtrack points) and see whether they are as you
expect them to be. Careful control of choice points can dramatically
reduce the memory footprint of a computation.

Results from efficiency tests are not usually portable between
different Prolog environments.  Prolog environment developers usually
try to ensure that performance of a particular construct does not
deteriorate in later versions, but sometimes an alternative construct
may become much faster. Make experiments in the specific version of
the Prolog environment that you eventually intend to run your program
in. Where experiments lead you to choose one construct over another,
document the reason for the choice by adding comments to the code.

%-----------------------------------------------------------------------
\guideline{Conduct experiments by writing separate small programs, not by mangling the main one.}
\label{dontmangle}

You will often have to experiment to pin down the exact behavior of a
poorly-documented built-in routine or to determine which of two
computations is more efficient.
Do this by writing separate small programs so that you will know
exactly what you are experimenting with.  Do not risk making erroneous
changes in a larger program in which you have invested considerable
effort.  There is always the risk that if you make experimental
changes, you will forget to undo them (or, worse, forget \emph{how} to
undo them).
Even if a code versioning system\footnote{Highly recommended:
see Guideline~\ref{guideline:use_a_revision_control_system}.}
makes it easy to roll back
experimental changes, \emph{microbenchmarks} are often more indicated
to assess the efficiency of alternative solutions.
In fact, when the experiment is conducted on a large program,
the changes frequently have effects that are harder to understand because
there is more code for them to interact with.

%-----------------------------------------------------------------------
\guideline{Look out for constructs that are almost always wrong.}

These include:
\begin{itemize}
\item
a cut at the end of the last clause of a predicate (exactly what
alternatives is it supposed to eliminate?);\footnote{Such
a cut can be legitimate, e.g., when a subgoal in the last clause
leaves choice points that need to be discarded (even though
the use of \texttt{once/1} is a better option).  In any case, a
predicate ending with a cut is so often a programming error that it
bears careful scrutiny, and if intentional, should be commented as
such (and \emph{still} scrutinized).}
\item
a {\tt repeat} not followed by a cut (when will it stop repeating?);
\item
\verb"append" with a one-element list as its first argument (use the
element and \verb"|" instead).
\end{itemize}

%=======================================================================
%\subsection{Reliability}

%-----------------------------------------------------------------------
\guideline{Isolate non-portable code.}

Suppose you are writing an SWI-Prolog program that passes some commands to Windows. You might be tempted to put calls such as
\begin{verbatim}
    ...,
    win_exec(X, normal),
    ...
\end{verbatim}
all through it.
Resist the temptation. Instead, define a predicate:
\begin{verbatim}
pass_command_to_windows(X) :-
    win_exec(X, normal).
\end{verbatim}
Elsewhere in your program, call \verb"pass_command_to_windows",
not \verb"win_exec".
That way, when you move to another Prolog system in which
\verb"win_exec" has a different name or works differently, you will
only need to change one line in your program.

%-----------------------------------------------------------------------
\guideline{Isolate ``magic numbers'' and other constants.}

If a number occurs more than once in your program, make it the
argument of a fact. Instead of writing
\begin{verbatim}
    X is 3.14159 * Y
\end{verbatim}
in one place and
\begin{verbatim}
    Z is 3.14159 * Q
\end{verbatim}
somewhere else, define a fact:
\begin{verbatim}
pi(3.14159).
\end{verbatim}
and write
\begin{verbatim}
    pi(P),
    X is P * Y
\end{verbatim}
and so forth.

This example may seem silly because the value of $\pi$ never changes,
but if you encode $3.14159$ in only one place, you don't have to worry
about mis-typing it elsewhere.

And if a number in your program really \emph{does} change ---a number
denoting an interest rate, or the maximum size of a file, or
something--- then it is very important to be able to update the program
by changing it in just one place, without having to check for other
occurrences of it.

It is common practice to define a predicate \verb"setting/2" to store
constants that are likely to be changed --- essentially, parameters
that are hard-coded into the program, such as:
\begin{verbatim}
setting(require_login, true).
setting(timeout_milliseconds, 500).
setting(output_directory, '/home/users/student42/project/output').
...
\end{verbatim}

%-----------------------------------------------------------------------
\guideline{Take the extra minute to \emph{prevent} errors rather than having to find them later.}

This is a basic rule of programming. Don't take the extra hour ---
just take the extra \emph{minute}.
In particular, think through loops. For example, given the predicate
\begin{verbatim}
count_up(10) :-
    !.
count_up(X) :-
    write(X),
    Y is X + 1,
    count_up(Y).
\end{verbatim}
what is the first number printed when you count up from $1$?  The last
number? What happens if you try to count up from $11$, or from $0.5$?
A minute's careful thought at the right time can save you an hour of
debugging on the computer.

%-----------------------------------------------------------------------
\guideline{Test code at its boundaries (limits).}

With \verb"count_up" above, the boundary is 10, so you should be
especially attentive to situations where the
argument is near 10 --- slightly below 10, or exactly 10, or slightly above it.

%-----------------------------------------------------------------------
\guideline{Test that each loop starts correctly, advances correctly,
and ends correctly.}

In \verb"count_up":
\begin{itemize}
\item What is the first number printed?
\item After printing a particular number (such as 5), what gets printed next?
\item What is the last number printed?
\end{itemize}
That's the essence of understanding any loop: you have to know where
it starts, how it advances from one value to the next, and where it
stops.

%-----------------------------------------------------------------------
\guideline{Test every predicate by failing back into it.}

It is not enough to try
\begin{verbatim}
?- count_up(5).
\end{verbatim}
and see what prints out. You must also try
\begin{verbatim}
?- count_up(5), fail.
\end{verbatim}
and see what happens when \verb"count_up" is forced to backtrack. This
will show you why the first clause of \verb"count_up" has to contain a
cut.

%-----------------------------------------------------------------------
\guideline{Do not waste time testing for errors that will be caught anyway.}

If you type
\begin{verbatim}
?- count_up(five).
\end{verbatim}
the program throws an arithmetic exception.
That's probably okay; the important thing is that you should know that
it will happen.

Do not burden your Prolog programs by testing the type of every
argument of every predicate.
Check only for errors that (1) are likely to occur, and (2) can be
handled by your program in some useful way.
For example, what happens if you execute this query?
\begin{verbatim}
?- count_up(What).
\end{verbatim}
The most common ``arguments of the wrong type'' are uninstantiated arguments.
Beware: they match anything!

The documentation for a predicate should make it clear which types of
arguments are expected and which patterns of argument instantiation
are supported. Where you think it is likely that incorrect argument
types might be supplied, then add a check to protect against
this. However, to avoid rechecking the arguments at each level of recursion,
consider adding a separate entry-point predicate that checks the
arguments before invoking the original predicate. The original
predicate will, of course, need to be renamed.

%-----------------------------------------------------------------------
\guideline{For anything beyond a proof-of-concept prototype, use testing systematically.}

Equip yourself with some infrastructure to conduct systematic
regression testing. Run tests frequently, at least whenever
substantial code changes are made.
Some Prolog systems provide excellent tools for unit testing.
For example, SWI-Prolog provides
\verb"plunit",\footnote{See \url{http://www.swi-prolog.org/pldoc/package/plunit.html}.} a tool that is meant to be portable to other systems as well.

%-----------------------------------------------------------------------
\guideline{In any error situation, make the program either correct the problem or crash (not just fail).}

In Prolog, \emph{failing} is not the same as \emph{crashing}. Crashing
here means calling the ISO Prolog standard built-in predicate
\verb"throw/1".  If the exception is not
caught by the program, the program will terminate with an error.

Failing is the same as answering a question ``no.''  ``No'' can be a
truthful and informative answer, and it is not the same as saying,
``Your question does not make sense,'' or ``I cannot compute the
answer.''  An example:
\begin{verbatim}
?- square_root(169, 13).
yes

?- square_root(169, 12).
no         % makes sense because 12 is not the square root of 169

?- square_root(georgia, X).
no         % misleading because the question is ill-formed
\end{verbatim}
In the last case the program should complain that \verb"georgia" is
not a number.

%-----------------------------------------------------------------------
\guideline{Make error messages informative.}

Whenever your program outputs an error message, that message must
actually explain what is wrong, and, if possible, display the data
that caused the problem. It should also indicate where the error was
detected (e.g., by giving the predicate name). A good example:
\begin{verbatim}
mysort/3: 'x(y,z,w)' cannot be sorted because it is not a list
\end{verbatim}
A bad example:
\begin{verbatim}
error: illegal data, wrong type
\end{verbatim}
That doesn't say \emph{what} data is the wrong type or where it was found.

Error reports in library files or modules should be related to the
exported predicate they originated from, not to some internal
predicate which the user doesn't know about.  To achieve this, it may
be necessary to add error handlers to some exported predicates just so
that they can handle internal deeply generated exceptions and map them
back to meaningful user-level descriptions.

%-----------------------------------------------------------------------
\guideline{Master the Prolog debugger; it is simple, powerful, and portable.}

The classic Prolog debugger allows you to step through a program. It
works the same way in virtually all compilers and is described in
\cite{ClocksinM03,CovingtonNV97} and other books.

If your Prolog environment supports graphical debugging then learn how
to use it.  The principles are the same as the standard command line
debugger, but much more information can be displayed on-screen at
once, and the program state is more readily visible.

%-----------------------------------------------------------------------
\guideline{Do not use {\tt write} for debugging.}

The easiest way to see what a program does is to sprinkle it with
\verb"print/1-2" output commands (when supported by your Prolog compiler),
so that you can see the values of the
variables as the computation proceeds. This is a good tactic in all
programming languages.

In Prolog, don't let backtracking confuse you. You might want to put a
unique number in each \verb"print/1-2" so that you can tell exactly which
one produced any particular line of output.

The advantage of using \verb"print/1-2" instead of \verb"write/1-2" is that
the output of \verb"print/1-2" can be tailored and, in particular,
appropriately abbreviated for debugging whereas \verb"write/1-2" insists
on showing you everything.

When the output from your program is not being displayed or where
debug output would be hard to see for some other reason, remember that
you can use \verb"print/2" to explicitly direct output to the terminal
or to a log file.

If you add tracing print commands to a program until you don't need
any more, you will find that you have too many to make sense of the
output.  What you need is selective tracing.  For example, in UNIX
the C \verb"syslog()" function has a priority argument with values
representing emergency, critical condition, error, warning, notice,
information, and debugging.  This allows the production of tracing
output to be controlled by how important it is.  That's often not
enough.  Part of your program may be suspect, and other parts not,
so you might want everything about one topic to be reported while
only errors should be reported elsewhere.

The answer is language-independent common practice.  Look
for a ``tracing'' or ``logging'' facility that lets you say ``here
is a message at this severity level about this topic with this
content'', where the topics and severities that get reported can
be selected at run time.  SWI-Prolog, for example, has
\verb"print_message/2", which offers severity level and
programmer-defined formatting, and \verb"debug/2", which offers
topic and format-string formatting.  If you don't find exactly what
you need, write your own logging library.

You should normally leave logging commands in your source code and
suppress their output by selecting no topics and only severe levels.
Do not assume that their presence is hurting performance until you
have measurements that prove it.  Use compiler options to remove
debugging and logging code.  If you are using your own logging
library, and your Prolog system supports term expansion or goal
expansion, you might be able to rewrite unwanted logging goals to
\verb"true".

%-----------------------------------------------------------------------
\guideline{Use a revision control system.}
\label{guideline:use_a_revision_control_system}

Every serious programmer can benefit from a revision control system
(version control system).  For the individual developer a revision
control system provides an audit trail of changes to a program; a way
to retrieve previous versions; and a way to manage different branches
of your code.  For multiple developers, a revision control system
makes collaborative work possible.

The lowest form of revision control system is to make a new backup
copy of the \emph{entire} project before every work session.  This is,
of course, wasteful of disk space.  Proper revision control systems
are available as local, client-server, and distributed systems.
Commonly used packages include CVS, Subversion, Mercurial, and Git.
There are numerous free and commercial implementations. Choose one
that suits your needs and use it.

As well as your Prolog programs, keep your test cases, documentation,
data files and associated scripts under revision control too. For
scientific work, repeatability is essential so make sure that you flag
the version used in a published work. Many revision control systems
have a web interface that is an ideal way to make your programs
publicly available.

\bigskip
\begin{minipage}{\columnwidth}
\begin{flushright}
{\small\sf
Be sure to comment out unused code instead of deleting it \\
and relying on version control to bring it back if necessary.\\
In no way document whether the new code was intended to supplement\\
or completely replace the old code, or whether the old code worked\\
at all, what was wrong with it, why it was replaced, etc.  Comment it\\
out with a lead \verb"/*" and trail \verb"*/" rather than a [\verb"%"]\\
on each line.  That way it might more easily be mistaken for live code\\
and partially maintained.\\}
--- \textsc{Roedy Green} \cite{Green96-10}
\end{flushright}
\end{minipage}

%=======================================================================
\section{Conclusion}
\label{sec:conclusion}
%=======================================================================

It is widely acknowledged that the adoption of coding standards is one
of the key factors in the success of a project:
it makes it possible to cut the software maintenance costs;
it enables effective collaboration between developers;
it may even prevent implosion of the project (due to poor readability,
lack of useful documentation, use of conflicting conventions
and interfaces \dots plus of course the friction
within the development team all this usually generates).

For a language like Prolog, coding standards are even more important,
due to the following factors:
\begin{enumerate}
\item
powerful (and complex) language features, such as multiple modes, dynamic code,
meta-programming facilities;
\item
lack of prescriptive typing and other machine-checkable declarations;
\item
substantial lack of sophisticated software development tools.
\end{enumerate}
Despite its potential benefits, a coherent and reasonably complete
set of coding guidelines for Prolog has, to the best of our
knowledge, never been published.  Moreover, no \emph{de facto} standard seems
to have emerged.  This can be partly explained by the fact that, also due to
the lack of a comprehensive language standard, the user community is
fragmented into sub-communities that are centered around individual Prolog
systems.

In this paper we have made a first step towards filling this gap.
We have highlighted those aspects of Prolog program development that
deserve particular attention and would benefit from a disciplined approach.
For each of those we have introduced a set of coding guidelines, illustrating
the underlying rationale. Where alternative
(or even conflicting) guidelines could achieve the desired effect,
their relative merits have been discussed.

In addition to guidelines that are valuable for any Prolog programmer,
this paper contains advice that may be obvious to the more expert.
This is intentional: as truly expert Prolog programmers are in very short
supply (logic programming courses have been dropped in many universities),
the availability and adoption of even simple coding guidelines
can improve the productivity of many projects.

Of course, a good degree of subjectiveness is unavoidable on these
matters. Nonetheless, we have explained the reasons why it is
important to pay attention to certain aspects of Prolog and provided
examples that may serve as a basis for further elaboration.
Such an elaboration is unavoidable: due to differences in project
purposes, environments and developer communities, a full-fledged coding
standard can only be established on a per-project basis.
We believe the present paper provides a useful starting point in this
respect.

\paragraph{Acknowledgment}

The authors gratefully thank all those who provided feedback
and suggestions to improve the paper, among which:
Abramo Ba\-gna\-ra,
Alan Baljeu,
Jan Burse,
Parker Jones,
G\"unter Kniesel,
Andrzej Lewandowski,
Naomi Lindenstrauss,
Jos\'e Morales,
Paulo Moura,
Robert Muetzelfeldt,
Ulrich Neumerkel,
Philip Smith,
Sylvain Soliman,
Michael Sperberg-McQueen,
Walter Wilson, and the anonymous reviewers.
The input provided by Andrzej Lewandowski and Paulo Moura was particularly
valuable.

%=======================================================================
%\bibliographystyle{acmtrans}
%\bibliography{plcoding}

% This was generated by BibTeX but now has been copied here for hand-editing.

\end{document}